\documentclass[aps,prd,twocolumn,superscriptaddress,showpacs,preprintnumbers,amsmath,amssymb]{revtex4}
\usepackage{graphicx}
\usepackage{dcolumn} 

\graphicspath{{ps}}

\begin{document}

\title{ \quad\\[1.0cm]  Measurement of
$D^+ \rightarrow K_S^0 K^+$ and $D^+_s \rightarrow K_S^0 \pi^+$
branching ratios
}
\affiliation{Budker Institute of Nuclear Physics, Novosibirsk}

\affiliation{University of Cincinnati, Cincinnati, Ohio 45221}

\affiliation{Justus-Liebig-Universit\"at Gie\ss{}en, Gie\ss{}en}
\affiliation{The Graduate University for Advanced Studies, Hayama}

\affiliation{Hanyang University, Seoul}
\affiliation{University of Hawaii, Honolulu, Hawaii 96822}
\affiliation{High Energy Accelerator Research Organization (KEK), Tsukuba}

\affiliation{Institute of High Energy Physics, Chinese Academy of Sciences, Beijing}
\affiliation{Institute of High Energy Physics, Vienna}
\affiliation{Institute of High Energy Physics, Protvino}

\affiliation{Institute for Theoretical and Experimental Physics, Moscow}
\affiliation{J. Stefan Institute, Ljubljana}
\affiliation{Kanagawa University, Yokohama}

\affiliation{Korea University, Seoul}

\affiliation{Kyungpook National University, Taegu}
\affiliation{\'Ecole Polytechnique F\'ed\'erale de Lausanne (EPFL), Lausanne}
\affiliation{Faculty of Mathematics and Physics, University of Ljubljana, Ljubljana}
\affiliation{University of Maribor, Maribor}
\affiliation{Max-Planck-Institut f\"ur Physik, M\"unchen}
\affiliation{University of Melbourne, School of Physics, Victoria 3010}
\affiliation{Nagoya University, Nagoya}

\affiliation{Nara Women's University, Nara}
\affiliation{National Central University, Chung-li}
\affiliation{National United University, Miao Li}
\affiliation{Department of Physics, National Taiwan University, Taipei}
\affiliation{H. Niewodniczanski Institute of Nuclear Physics, Krakow}
\affiliation{Nippon Dental University, Niigata}
\affiliation{Niigata University, Niigata}
\affiliation{University of Nova Gorica, Nova Gorica}
\affiliation{Novosibirsk State University, Novosibirsk}
\affiliation{Osaka City University, Osaka}

\affiliation{Panjab University, Chandigarh}

\affiliation{University of Science and Technology of China, Hefei}
\affiliation{Seoul National University, Seoul}
\affiliation{Shinshu University, Nagano}
\affiliation{Sungkyunkwan University, Suwon}
\affiliation{School of Physics, University of Sydney, NSW 2006}
\affiliation{Tata Institute of Fundamental Research, Mumbai}
\affiliation{Excellence Cluster Universe, Technische Universit\"at M\"unchen, Garching}

\affiliation{Tohoku Gakuin University, Tagajo}

\affiliation{Department of Physics, University of Tokyo, Tokyo}

\affiliation{Tokyo Metropolitan University, Tokyo}
\affiliation{Tokyo University of Agriculture and Technology, Tokyo}

\affiliation{IPNAS, Virginia Polytechnic Institute and State University, Blacksburg, Virginia 24061}
\affiliation{Yonsei University, Seoul}
  \author{E.~Won}\affiliation{Korea University, Seoul} 
  \author{B.~R.~Ko}\affiliation{Korea University, Seoul} 

  \author{H.~Aihara}\affiliation{Department of Physics, University of Tokyo, Tokyo} 
  \author{K.~Arinstein}\affiliation{Budker Institute of Nuclear Physics, Novosibirsk}\affiliation{Novosibirsk State University, Novosibirsk} 

  \author{V.~Aulchenko}\affiliation{Budker Institute of Nuclear Physics, Novosibirsk}\affiliation{Novosibirsk State University, Novosibirsk} 
  \author{T.~Aushev}\affiliation{\'Ecole Polytechnique F\'ed\'erale de Lausanne (EPFL), Lausanne}\affiliation{Institute for Theoretical and Experimental Physics, Moscow} 

  \author{A.~M.~Bakich}\affiliation{School of Physics, University of Sydney, NSW 2006} 
  \author{V.~Balagura}\affiliation{Institute for Theoretical and Experimental Physics, Moscow} 

  \author{E.~Barberio}\affiliation{University of Melbourne, School of Physics, Victoria 3010} 
  \author{A.~Bay}\affiliation{\'Ecole Polytechnique F\'ed\'erale de Lausanne (EPFL), Lausanne} 

% ...
\collaboration{The Belle Collaboration}

\begin{abstract}
We report an improved measurement of 
$D^+ \rightarrow K_S^0 K^+$ and $D^+_s\rightarrow K_S^0 \pi^+$ branching ratios 
using 605 fb$^{-1}$ of data
collected with the Belle detector at the KEKB asymmetric-energy
$e^+ e^-$ collider. 
The measured branching ratios with respect to the Cabibbo-favored modes are
$\mathcal{B}(D^+ \rightarrow K_S^0 K^+)/\mathcal{B}(D^+ \rightarrow K_S^0 \pi^+)$ = 0.1899$\pm$0.0011$\pm$0.0022 
and 
$\mathcal{B}(D^+_s \rightarrow K_S^0 \pi^+)/\mathcal{B}(D^+_s \rightarrow K_S^0 K^+)$ = 0.0803$\pm$0.0024$\pm$0.0019 
where the first uncertainties are statistical and the second are systematic.
\end{abstract}


\maketitle

\tighten

{\renewcommand{\thefootnote}{\fnsymbol{footnote}}}
\setcounter{footnote}{0}

 Decays of charmed mesons play an important role 
in understanding the sources of SU(3) flavor symmetry breaking~\cite{ref:rosner}. 
Such a breaking can originate from
strong final-state interactions or interference between amplitudes with 
the same final state.
In particular,
$D^+ \rightarrow \overline{K}^0 K^+$ 
and $D^+_s \rightarrow K^0 \pi^+$~\cite{ref:cc}
are Cabibbo-suppressed (CS) decays that involve color-favored tree, 
annihilation, and penguin diagrams.  For $D^+$ decays, the branching ratio 
$\mathcal{B}(D^+ \rightarrow \overline{K}^0 K^+)/\mathcal{B}(D^+ \rightarrow 
\overline{K}^0 \pi^+)$ deviates from the naive $\tan^2{\theta_C}$ 
expectation~\cite{ref:pdg}, due to the destructive interference 
between color-favored and color-suppressed amplitudes
in $D^+ \rightarrow \overline{K}^0 \pi^+$~\cite{ref:guberina}.
However, converting experimental measurements of $D$ decays that include
$K_S^0$ branching ratios to those involving $K^0$ or $\overline{K}^0$ is 
not straightforward due to the interference between the doubly Cabibbo-suppressed 
(DCS) and Cabibbo-favored (CF) decay modes where
the interference phase is unknown~\cite{ref:bigi,ref:cleo_dksk}.
In $D^+_s$ decays to $\overline{K}^0 K^+$ and $K^0 \pi^+$ final states, the 
ratio of the CS decay to the corresponding CF decay may be larger than 
$\tan^2{\theta_C}$,
since the tree diagram for 
$D^+_s \rightarrow \overline{K}^0 K^+$ 
is CF but color-suppressed.
Precise measurements of branching ratios for CS and CF charmed
meson decay modes can thus improve the understanding of the 
underlying dynamics of these decays.
In this paper, we report improved measurements of the 
$D^+ \rightarrow K_S^0 K^+$ and $D^+_s \rightarrow K_S^0 \pi^+$
branching ratios with respect to the corresponding CF modes,
$D^+ \rightarrow K_S^0 \pi^+$ and $D^+_s \rightarrow K_S^0 K^+$,
respectively.

The results are based on a data sample of 605 fb$^{-1}$ recorded at the
$\Upsilon$(4S) resonance with the Belle detector at
the KEKB asymmetric-energy $e^+e^-$  collider~\cite{ref:kekb}.
An additional data sample with about 10\% of this 
integrated luminosity recorded 60 MeV
below the $\Upsilon$(4S) was used for the
optimization of the selection criteria (off-resonance
sample).
The Belle detector is a large-solid-angle magnetic spectrometer that consists
of a silicon vertex detector (SVD), a 50-layer central drift chamber (CDC), an array of
aerogel threshold Cherenkov counters (ACC), a barrel-like arrangement of time-of-flight
scintillation counters (TOF), and an electromagnetic calorimeter comprised of CsI(Tl) 
crystals located inside a superconducting solenoid coil that provides a 1.5 T
magnetic field. An iron flux return located outside the coil is instrumented
to detect $K^0_L$ mesons and to identify muons. The detector is described in detail
elsewhere~\cite{ref:belle}.

We require that the charged tracks originate from the vicinity of
the interaction point (IP)
with the impact parameters in the beam direction ($z$-axis) 
and perpendicular to it of less than 4 cm and 2 cm, respectively. All charged
tracks except those originating from $K_S^0$ decays
are required to have at least two associated hits in the SVD,
both in the $z$ and radial directions, to assure good
spatial resolution on the $D$ mesons' decay vertices. 
Charged tracks are identified as pions or kaons by requiring the
ratio of particle identification likelihoods,
$\mathcal{L}_K/(\mathcal{L}_K+\mathcal{L}_\pi)$, constructed using information
from the CDC, TOF, and ACC,
be larger or smaller than 0.6, respectively. 
For both kaons and pions,
the efficiencies and misidentification probabilities are 86\% and 10\%, respectively. 

Pairs of oppositely charged tracks that have an invariant mass within 
30 MeV/$c^2$ of the nominal $K_S^0$ mass are used to reconstruct 
$K_S^0 \rightarrow \pi^+ \pi^-$ decays. The distance of the closest approach 
of the candidate charged tracks to the IP
in the plane perpendicular to the
$z$ axis is required to be larger than 0.02 cm for high-momentum
($>$ 1.5 GeV/$c$) $K_S^0$ candidates and 0.03 cm for those with
momentum less than 1.5 GeV/$c$. The $\pi^+\pi^-$ vertex is required
to be displaced from the IP by a minimum transverse distance of
0.22 cm for high-momentum candidates and 0.08 cm for the remaining
candidates. The mismatch in the $z$ direction at the $K_S^0$ vertex
point for the $\pi^+\pi^-$  tracks must be less than 2.4 cm
for high-momentum candidates and 1.8 cm for the remaining candidates.
The direction of the pion pair momentum must also agree with the direction
of the vertex point from the IP to within 0.03 rad for high-momentum 
candidates and to within 0.1 rad for the remaining candidates.

These two sets of criteria in two different momentum ranges are implemented 
to maximize $\mathcal{N}_S/\sqrt{\mathcal{N}_S+\mathcal{N}_B}$, 
where $\mathcal{N}_S$ and $\mathcal{N}_B$ are the number of signal 
$K_S^0$'s and the number of combinatorial background events, respectively.
Finally, the $\pi^+ \pi^-$ pair forming a $K_S^0$ candidate  is 
required to have an invariant 
mass within $\pm$9 MeV/$c^2$ of the nominal $K_S^0$ mass~\cite{ref:pdg}. 

$D^+$ and $D^+_s$ candidates are reconstructed using a $K_S^0$ candidate and
either a charged pion or kaon candidate.
The decay vertex is formed by fitting the $K_S^0$ and the track 
to a common vertex and requiring a confidence level greater than 0.1\%.  In order to remove 
peaking backgrounds from the $D^+_{(s)} \rightarrow \pi^+\pi^+\pi^-$ and $K^+\pi^+\pi^-$ decay 
modes, we compute the reduced $\chi^2$ of the vertex assuming that two pions
from the $K_S^0$ and the charm daughter track arise from a single vertex. We require 
the reduced $\chi^2$ to be greater than 10. 

To remove $D^+$ and $D^+_s$ mesons produced in $B$ meson decays, 
we require the charmed meson momentum calculated in
the center-of-mass frame to be greater than 2.6 GeV/$c$.
At this stage, reconstruction efficiencies are 16.6\% for the $D^+$ and 
18.0\% for the $D^+_s$ in the $K_S^0 K^+$ final state, 
and 20.6\% for the $D^+$ and 22.4\% for the $D^+_s$ in the $K_S^0 \pi^+$ 
final state. 

Highly asymmetrical $K_S^0 h^+$ pairs that have 
invariant mass close to the $D^+_{(s)}$ mass region are
more likely to be background than signal. 
The asymmetry, $\mathcal{A}$ $\equiv$ $|p_{K_S^0} - p_{h^+}|/|p_{K_S^0} + p_{h^+}|$
, where each momenta is calculated in the laboratory 
frame and $h^+$ refers to either a $K^+$ or $\pi^+$,
is used to reject background candidates.
The $\mathcal{A}$ requirement is optimized in both CS modes 
by maximizing $\mathcal{N}_S/\sigma_{\mathcal{N}_S}$, where $\mathcal{N}_S$ 
is the signal yield and $\sigma_{\mathcal{N}_S}$ is the statistical 
uncertainty in $\mathcal{N}_S$ from the fit to the off-resonance data sample.
The asymmetry is required to be less than 0.6 for both decay modes.
After this final requirement, we find 10\% and 35\% improvements
in $\mathcal{N}_S/\sigma_{\mathcal{N}_S}$ for CS decay modes of the $D^+$ and 
$D^+_s$, respectively.

 Since there are differences in the mass distributions between the data and 
Monte Carlo (MC) simulated~\cite{ref:bellemc} samples, we tune
the large MC samples of generic continuum and $B\bar{B}$ decays,
intended mainly for the accurate parameterization of the peaking background
under the signal. This background is a consequence of particle 
misidentification and will be discussed in more detail later.  
The tuning procedure is as follows:
the $\pi^+$ ($K^+$) momentum scale and
resolution are tuned with the $D^0 \rightarrow \pi^+\pi^-$ 
($D^0 \rightarrow K^+K^-$) data sample. 
For the $K_S^0$ momentum scale and resolution tuning,
the $D^+ \rightarrow K_S^0 \pi^+$ data sample is used.
The tuning method is validated by comparing
simulated and real data in the $K_S^0 K^+$ final state.
The four signal decay modes are simulated and results of the tuning 
are also applied to them.

In the branching ratio measurements, there is  a peaking background 
due to particle misidentification.  In the $D^+_s \rightarrow K_S^0 K^+$ mass 
region, there is a peaking structure from $D^+ \rightarrow K_S^0 \pi^+$ decays 
when a $\pi^+$ is misidentified as a $K^+$.  A similar peaking structure in 
the $D^+ \rightarrow K_S^0 \pi^+$ mass region appears due to 
misidentification in $D^+_s \rightarrow K_S^0 K^+$ decays.
The shapes and the yields of these peaking backgrounds
are obtained from the tuned simulation samples
and are used as the probability density functions (PDF) 
for the peaking backgrounds. 
The simulated shape and normalization of the peaking backgrounds are checked
by comparing the invariant mass distributions of 
selected $K_S^0K^+$ ($K_S^0 \pi^+$) pairs with the $K^+$~($\pi^+$) mass 
assignment changed to a $\pi^+$~($K^+$) mass assignment. The 
comparison shows that the simulated peaking background of the tuned 
sample correctly describes these components and that 
misidentification is indeed the only contribution above the structureless 
combinatorial background. Uncertainties in the misidentification 
probabilities are considered as a source of systematic uncertainty.

\begin{figure}[htbp]
\includegraphics[height=0.47\textwidth,width=0.5\textwidth]{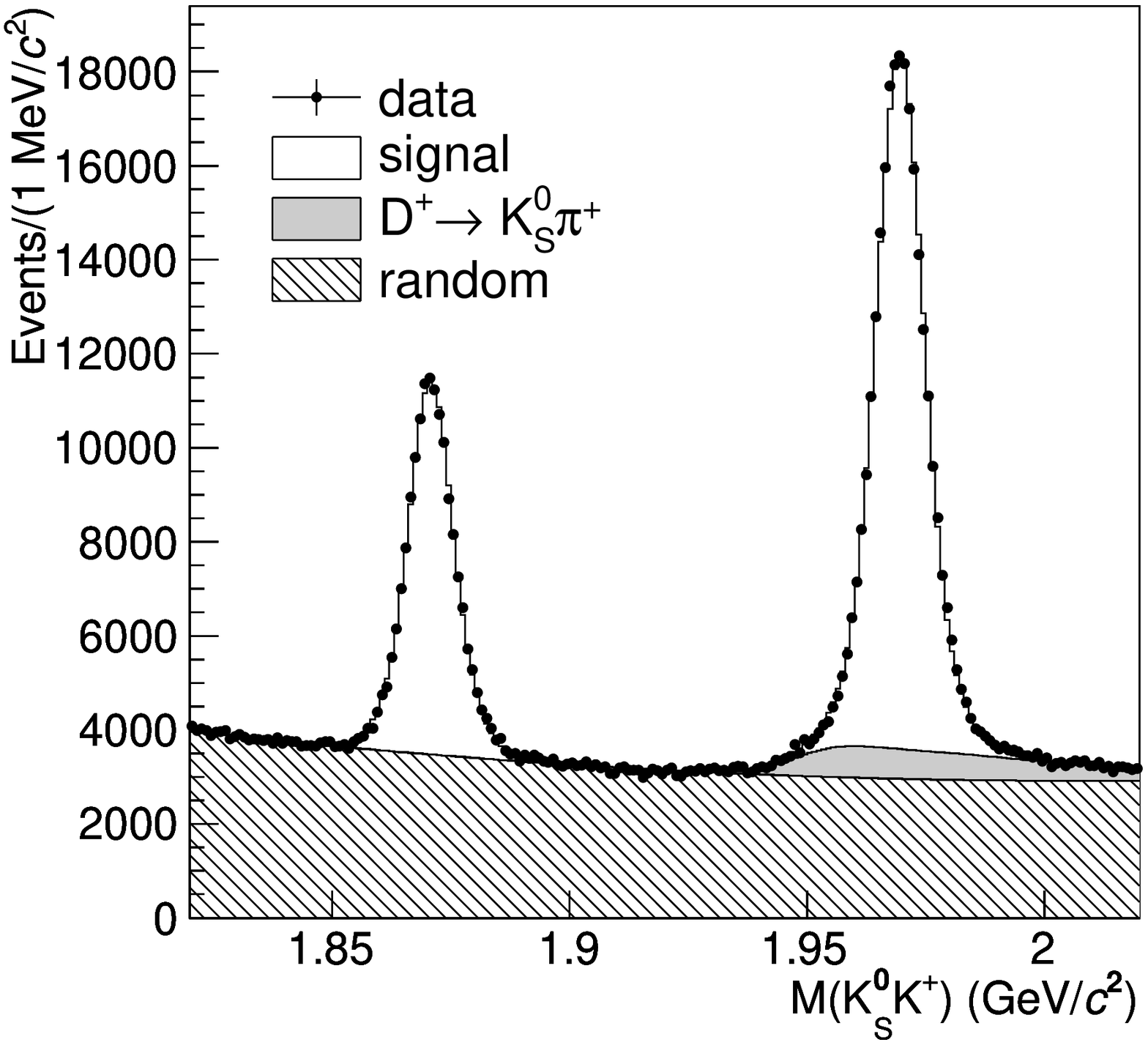}
\caption{Invariant mass distribution of 
selected
$K_S^0 K^+$ pairs. 
Points with error bars
show the data and histograms show the results of the fits described in
the text. Signal, $D^+ \rightarrow K_S^0 \pi^+$ background, and random
combinatorial background components are also shown.
}
\label{fig:data_ksk}
\end{figure}

\begin{figure}[htbp]
\includegraphics[height=0.47\textwidth,width=0.5\textwidth]{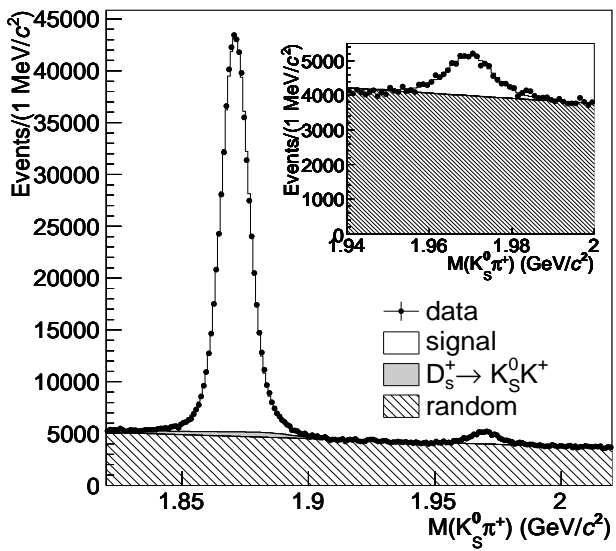}
\caption{Invariant mass distribution of 
selected
$K_S^0 \pi^+$ pairs. 
Points with error bars
show the data and histograms show the results of the fits described in
the text. Signal, $D^+_s \rightarrow K_S^0 K^+$ background, and random
combinatorial background components are also shown.
The inset is an enlarged view of the $D^+_s$ region.
}
\label{fig:data_kspi}
\end{figure}

The $K_S^0 K^+$ and $K_S^0 \pi^+$ invariant mass distributions after the final
selections are shown in Figs.~\ref{fig:data_ksk} and \ref{fig:data_kspi} 
together with the signal and background parameterizations. 
Clear signals for CF and CS decays are observed
in both distributions. The $K_S^0 K^+$ and $K_S^0 \pi^+$ invariant mass distributions are
fitted using a binned maximum likelihood method. In all cases the signal 
PDF is parameterized using two Gaussians with 
a common mean value. For $D^+_s \rightarrow K_S^0 \pi^+$,  we
fix the ratio of widths and the fractional yields in the two Gaussians 
because of the low statistics. 
The values of the ratio and the fraction 
of the broader Gaussian are obtained from 
the fit to the $D^+ \rightarrow K_S^0 \pi^+$ mode 
and are consistent with the results of fits to MC simulated signal. 
The reduced $\chi^2$ values of the fits are 1.8 and 2.3 for the $K_S^0 K^+$ 
and $K_S^0 \pi^+$ final states, respectively.
The normalization of the mass distributions of the misidentified
$K/\pi$ backgrounds are fixed to the values obtained from tuned 
simulation samples. 
Combinatorial background PDFs are parameterized using second and 
first-order polynomials for the $K_S^0 K^+$ and $K_S^0 \pi^+$
final states, respectively.
All the fit parameters are allowed to float 
except for the $D^+_s \rightarrow K^0_S \pi^+$ signal PDF parameters 
and the yield and the normalization of the misidentified backgrounds.
Table~\ref{table:fit_result} summarizes the extracted signal
yields from the fits to data and corresponding signal efficiencies 
from the simulated signal samples where final-state
radiation has been included~\cite{ref:photos}.

\begin{table}[htb]
\caption{ The extracted signal yields from the fits to data and 
corresponding signal efficiencies ($\epsilon$) from the simulated 
events of signal modes. The uncertainties are statistical only.
}
\label{table:fit_result}
\begin{ruledtabular}
\begin{tabular}
 {@{\hspace{0.5cm}}c@{\hspace{0.5cm}} 
  @{\hspace{0.5cm}}c@{\hspace{0.5cm}}  @{\hspace{0.5cm}}c@{\hspace{0.5cm}}}
Decay modes & Yields  & $\epsilon$ (\%)\\
\hline
$D^+   \rightarrow K_S^0 K^+ $  & 100855$\pm$561  & 12.59$\pm$0.01 \\
$D^+_s \rightarrow K_S^0 K^+ $  & 204093$\pm$768  & 13.53$\pm$0.01 \\
$D^+   \rightarrow K_S^0 \pi^+$ & 566285$\pm$1162 & 14.19$\pm$0.01 \\
$D^+_s \rightarrow K_S^0 \pi^+$ &  17583$\pm$481  & 15.35$\pm$0.01 \\
\end{tabular}
\end{ruledtabular}
\end{table}

Various contributions to the systematic uncertainties for the branching
ratio measurements are summarized in Table~\ref{table:sys}.  
Several sources of systematic uncertainty are reduced in ratio 
measurements due to the similar kinematics of CF and CS decays. Such 
sources include the tracking and asymmetry variable efficiency 
differences between simulated data and real data. 
However, the systematic uncertainty due to particle identification 
efficiency does not cancel. 
The particle identification efficiency 
differences between real data and simulated events are estimated independently 
using the decay $D^{*+} \rightarrow D^0 \pi^+$ followed by
$D^0 \rightarrow K^- \pi^+$ 
and corrections from this estimate
are applied to signal efficiencies in Table~\ref{table:fit_result}.
These corrections 
are 1.000$\pm$0.007 and 0.946$\pm$0.005 for $K^+$ and $\pi^+$
candidates, respectively.
Uncertainties in the particle identification 
corrections are included in the systematics estimate and are found to be 
0.90\% of the measured ratios. 
In order to validate the entire analysis procedure, we fit large numbers of 
simulated samples of generic continuum and $B\bar{B}$ decays,
and find no bias
in the procedure within the statistical uncertainties of our measurements.
We refit the data with various histogram binnings,
different fit intervals, 
and different combinatorial background PDFs.
We also refit the data in the $D^+$ and $D^+_s$ samples separately.
We estimate 0.74\% and 2.00\% of the measured ratios as the systematic 
uncertainties due to variations in fit methods for the $D^+$ and $D^+_s$ 
modes, respectively.  Particle identification and the associated normalizations
of the $K/\pi$ misidentified background yields in fits 
are also estimated using the measured misidentification rates 
and found to be 0.16\% and
0.62\% of the measured ratio for the $D^+$ and $D^+_s$ modes, respectively.
Finally, systematic effects due to the extra constraints in 
the $D^+_s \rightarrow
K_S^0\pi^+$ signal PDF are estimated by
refitting the data allowing the fixed parameters to change within their 
one standard deviation uncertainties.  This gives a negligible effect in 
$D^+$ decays but there is a systematic effect corresponding to 0.37\% of the
measured ratio in $D^+_s$ decay modes.
Table~\ref{table:sys} summarizes the systematic uncertainties in the 
branching ratio measurements.

\begin{table}[htb]
\caption{ Relative systematic uncertainties in percent,
where $\sigma_{R(D^+)}$ and $\sigma_{R(D^+_s)}$ are 
systematic uncertainties for branching ratios of
$D^+$ and $D^+_s$ decays. Sources include particle identification (PID),
fit methods, peaking background, and the $D^+_s$ signal PDF.
}
\label{table:sys}
\begin{ruledtabular}
\begin{tabular}
 {@{\hspace{0.5cm}}l@{\hspace{0.5cm}}  
  @{\hspace{0.5cm}}c@{\hspace{0.5cm}}  @{\hspace{0.5cm}}c@{\hspace{0.5cm}}}
Source &$\sigma_{R(D^+)}$ (\%) & $\sigma_{R(D^+_s)}$ (\%)\\
\hline
PID                     & 0.90  & 0.90 \\
Fit methods             & 0.74  & 2.00 \\
Peaking background & 0.16  & 0.62 \\
$D^+_s$ signal PDF      &  -    & 0.37 \\
\hline
Total & 1.18 & 2.31 \\
\end{tabular}
\end{ruledtabular}
\end{table}

 With the signal efficiencies and the corrections due to particle 
identification efficiency differences, we find the branching ratios to be  
\begin{eqnarray}
R(D^+)&\equiv&
\frac{\mathcal{B}(D^+ \rightarrow K_S^0 K^+)}{\mathcal{B}(D^+ \rightarrow K_S^0 \pi^+)}
\nonumber \\
&=& 0.1899\pm0.0011\pm0.0022,
\nonumber \\
R(D^+_s)&\equiv&
\frac{\mathcal{B}(D^+_s \rightarrow K_S^0 \pi^+)}{\mathcal{B}(D^+_s \rightarrow K_S^0 K^+)}
\nonumber  \\
&=& 0.0803\pm0.0024\pm0.0019 
\nonumber 
\end{eqnarray}
where the first uncertainties are  statistical and the second are systematic.
These are the most precise measurements to date and are compared to
the present world average values in Table~\ref{table:comparison}.
Our measurement of $R(D^+)$ is 
in good agreement with previous measurements~\cite{ref:pdg} 
and is larger than the
naive expectation of $\tan^2{\theta_C}$,
consistent with the expected destructive interference effect mentioned
earlier.  For $D^+_s$ decays, there is no such interference and 
$R(D^+_s)$ is found to be greater than 
$\tan^2{\theta_C}$ 
by more than eight standard deviations,
consistent with previous measurements~\cite{ref:pdg}.
This large deviation may be due to the 
color-suppression of the main $D^+_s \rightarrow K_S^0 K^+$ amplitude.

\begin{table}[htb]
\caption{Branching ratios for the $D^+$ and the $D^+_s$,  
and comparisons with previous measurements. The uncertainties shown combine 
the statistical and systematic uncertainties of our 
results.
}
\label{table:comparison}
\begin{ruledtabular}
\begin{tabular}
 {@{\hspace{0.5cm}}c@{\hspace{0.5cm}}  
  @{\hspace{0.5cm}}c@{\hspace{0.5cm}}  @{\hspace{0.5cm}}c@{\hspace{0.5cm}}}
Branching          & Belle      & World-  \\
ratio              & exp.       & average~\cite{ref:pdg} \\ 
\hline
$R(D^+)$ & (19.0$\pm$0.2)\%  & (20.6$\pm$1.4)\%\\
$R(D^+_s)$ & ( 8.0$\pm$0.3)\%  & ( 8.4$\pm$0.9)\%\\
\end{tabular}
\end{ruledtabular}
\end{table}

 To conclude, using 605 fb$^{-1}$ of data collected with the Belle detector 
at the KEKB asymmetric-energy $e^+e^-$ collider 
we have measured the
$D^+ \rightarrow K_S^0 K^+$ and 
$D^+_s \rightarrow K_S^0 \pi^+$ branching ratios 
with respect to the corresponding Cabibbo-favored modes. The results are
$\mathcal{B}(D^+ \rightarrow K_S^0 K^+)/\mathcal{B}(D^+ \rightarrow K_S^0 \pi^+)$ = 0.1899$\pm$0.0011$\pm$0.0022 
and 
$\mathcal{B}(D^+_s \rightarrow K_S^0 \pi^+)/\mathcal{B}(D^+_s \rightarrow K_S^0 K^+)$ = 0.0803$\pm$0.0024$\pm$ 0.0019, 
where the first uncertainties are statistical and the second are systematic. 
Using the world average values of CF decay rates~\cite{ref:pdg}, we obtain 
the branching fractions 
$\mathcal{B}(D^+ \rightarrow K_S^0 K^+)$ = (2.75$\pm$0.08)$\times$10$^{-3}$ 
and 
$\mathcal{B}(D^+_s \rightarrow K_S^0 \pi^+)$ = (1.20$\pm$0.09)$\times$10$^{-3}$ 
where the uncertainties are the sum in quadrature of statistical and 
systematic errors.
These are consistent with the present world averages~\cite{ref:pdg} 
and are the most precise measurements to date. The ratio 
$\mathcal{B}(D^+ \rightarrow K_S^0 K^+)$/$\mathcal{B}(D^+_s \rightarrow K_S^0 \pi^+)$ = 2.29$\pm$0.18  
may be due to SU(3) flavor breaking and/or different 
final-state interactions in $D^+$ and $D^+_s$ decays.

We thank the KEKB group for the excellent operation of the
accelerator, the KEK cryogenics group for the efficient
operation of the solenoid, and the KEK computer group and
the National Institute of Informatics for valuable computing
and SINET3 network support.  We acknowledge support from
the Ministry of Education, Culture, Sports, Science, and
Technology (MEXT) of Japan, the Japan Society for the 
Promotion of Science (JSPS), and the Tau-Lepton Physics 
Research Center of Nagoya University; 
the Australian Research Council and the Australian 
Department of Industry, Innovation, Science and Research;
the National Natural Science Foundation of China under
contract No.~10575109, 10775142, 10875115 and 10825524; 
the Department of Science and Technology of India; 
the BK21 and WCU program of the Ministry Education Science and
Technology, the CHEP SRC program and Basic Research program (grant No.
R01-2008-000-10477-0) of the Korea Science and Engineering Foundation,
Korea Research Foundation (KRF-2008-313-C00177),
and the Korea Institute of Science and Technology Information;
the Polish Ministry of Science and Higher Education;
the Ministry of Education and Science of the Russian
Federation and the Russian Federal Agency for Atomic Energy;
the Slovenian Research Agency;  the Swiss
National Science Foundation; the National Science Council
and the Ministry of Education of Taiwan; and the U.S.\
Department of Energy.
This work is supported by a Grant-in-Aid from MEXT for 
Science Research in a Priority Area (``New Development of 
Flavor Physics''), and from JSPS for Creative Scientific 
Research (``Evolution of Tau-lepton Physics'').


\begin{thebibliography}{99}
\bibitem{ref:cc}
Throughout this paper, 
the inclusion of the charge-conjugate decay mode is implied
unless otherwise stated.
\end{thebibliography}
\end{document}